\documentclass[iop]{emulateapj}

\usepackage{apjfonts}
\usepackage{wrapfig}
\usepackage{amsmath}

\newcommand{\lir}{$L_{\rm IR}$}
\newcommand{\lagn}{$L_{\rm AGN}$}

\newcommand{\ergs}{erg s$^{-1}$}

\newcommand{\im}{\item}

\newcommand{\average}[1]{\ensuremath{\langle#1\rangle} }

\newcommand{\msun}{\ensuremath{M_{\sun}}}

\newcommand {\apgt} {\ {\raise-.5ex\hbox{$\buildrel>\over\sim$}}\ }
\newcommand {\aplt} {\ {\raise-.5ex\hbox{$\buildrel<\over\sim$}}\ }

\newcommand{\ben}{\begin{enumerate}}
\newcommand{\een}{\end{enumerate}}
\newcommand{\bit}{\begin{itemize}}
\newcommand{\eit}{\end{itemize}}
\newcommand{\beq}{\begin{equation}}
\newcommand{\eeq}{\begin{equation}}

\begin{document}

\title{Black hole variability and the star formation-AGN connection:\\ Do all star-forming galaxies host an AGN?}

\shorttitle{DO ALL STAR-FORMING GALAXIES HOST AN AGN?}
\shortauthors{HICKOX ET AL.}
\author{Ryan C.\ Hickox\altaffilmark{1}}
\author{James R.\ Mullaney\altaffilmark{2,3}}
\author{David M.\ Alexander\altaffilmark{3}}
\author{Chien-Ting J.\ Chen\altaffilmark{1}}
\author{Francesca M.\ Civano\altaffilmark{1,4}}
\author{Andy D.\ Goulding\altaffilmark{4}}
\author{Kevin N.\ Hainline\altaffilmark{1}}


\altaffiltext{1}{Department of Physics and Astronomy, Dartmouth College, 6127 Wilder Laboratory, Hanover, NH 03755; ryan.c.hickox@dartmouth.edu.}
\altaffiltext{2}{Department of Physics \& Astronomy, University of Sheffield, Sheffield, S3 7RH, United Kingdom}
\altaffiltext{3}{Department of Physics, Durham University, South Road, Durham, DH1 3LE, United Kingdom}
\altaffiltext{4}{Harvard-Smithsonian Center for Astrophysics, 60 Garden Street, Cambridge, MA 02138}

\slugcomment{Accepted for publication in The Astrophysical Journal}

\begin{abstract}
We investigate the effect of active galactic nucleus (AGN) variability
on the observed connection between star formation and black hole
accretion in extragalactic surveys.  Recent studies have reported
relatively weak correlations between observed AGN luminosities and the
properties of AGN hosts, which has been interpreted to imply that
there is no direct connection between AGN activity and star formation.
However, AGNs may be expected to vary significantly on a wide range of
timescales (from hours to Myr) that are far shorter than the typical
timescale for star formation ($\gtrsim$\,100 Myr). This variability
can have important consequences for observed correlations. We present
a simple model in which {\em all} star-forming galaxies host an AGN when
averaged over $\sim$\,100 Myr timescales, with long-term average AGN
accretion rates that are perfectly correlated with the star formation
rate (SFR). We show that reasonable prescriptions for AGN variability
reproduce the observed weak correlations between SFR and $L_{\rm AGN}$
in typical AGN host galaxies, as well as the general trends in the
observed AGN luminosity functions,  merger fractions, and
measurements of the average AGN luminosity as a function of SFR. These
results imply there may be a tight connection between AGN activity and
SFR over galaxy evolution timescales, and that the apparent similarities in rest-frame colors, merger
rates, and clustering of AGNs compared to ``inactive'' galaxies
may be due primarily to AGN variability. The results provide
motivation for future deep, wide extragalactic surveys that can
measure the {\em distribution} of AGN accretion rates as a function of
SFR.

\end{abstract}


\section{Introduction}
\label{sec:intro}

There has been a great deal of recent research activity investigating
the connection between supermassive black hole (BH) accretion and star
formation (SF) in galaxies (for a review see \S\,2--3 of
\citealt{alex12bh}). This work is largely motivated by observed
correlations between BHs and galaxy properties \citep[e.g.,][]{mago98,
  gebh00, ferr00, gult09msigma}, the remarkable similarity between the
global histories of SF and BH accretion \citep[e.g.,][]{boyl98qsosf,
  silv08xlf, aird10xlf, merl13agn}, and the observed tendency of
luminous active galactic nuclei (AGNs) to reside in star-forming hosts
\citep[e.g.,][]{lutz08qsosf, bonf11qsosf}. Furthermore, a number of
theoretical models predict that SF and BH growth should be closely
linked in galaxies, driven by a common supply of cold gas
\citep[e.g.,][]{dima05qso, hopk06merge, some08bhev, hopk10bhgas,
  angl13bhsim} and with energy released by the AGN
  potentially triggering new star formation
  \citep[e.g.,][]{naya12agnsf, zubo13agnsb, naya13feed}.

Contrary to these expectations, recent observational studies have
found relatively {\em weak} correlations between BH accretion and
SF. While some studies have reported a strong link between SF and AGN
activity for high-luminosity AGNs \citep[e.g.,][]{lutz08qsosf,
  bonf11qsosf}, at lower luminosities these correlations appear
relatively weak or absent. Furthermore, typical AGN hosts have star
formation rates (SFRs) characteristic of normal star-forming galaxies
at a given redshift \citep[e.g.][]{mull12agnsf}, and
moderate-luminosity AGNs exhibit similar SFRs to lower-luminosity
sources \citep[e.g.,][]{shao10agnsf, rosa12agnsf, harr12agnsf},
suggesting little connection between SF and BH growth in these
systems.

Similar conclusions can be drawn from the properties of AGN host
galaxies and dark matter halos. Moderate-luminosity AGNs appear to be
found in galaxies with a wide range of rest-frame optical or near-IR
colors and not simply in blue star-forming galaxies
\citep[e.g.,][]{nand07host, geor08agn, hick09corr, silv09xcosmos_env,
  rosa13xgal, goul13agn_aph}. AGN host galaxies have colors indistinguishable from
those of ``normal'' galaxies of similar mass and redshift
\citep[e.g.,][]{xue10xhost, card10xhost, bong12xgal, hain12agn,
  rosa13xgal}, although they do appear to show slightly stronger and
more centrally-concentrated SF \citep{sant12agnsf, rosa13agnsf,
  lama13agnsf, hick13seyf}. Likewise, the spatial clustering and thus
dark matter halo masses \citep[e.g.,][]{hick09corr, coil09xclust,
  dono10clust, li06agnclust} and morphologies
\citep[e.g.,][]{cist11agn, scha11xmorph, civa12ccosmos, koce12xmorph}
of moderate-luminosity AGNs are generally consistent with those for
``normal'' galaxies with similar mass, color, and redshift.

These results have frustrated many attempts to uncover the galaxy
properties responsible for ``triggering'' BH activity, leading some
authors to conclude that there is little, if any, connection between
the growth rate of BHs and SF in their hosts, at least in
moderate-luminosity systems \citep[e.g.,][]{shao10agnsf}. However,
other studies have suggested a {\em strong} correlation between BH
growth and SF.  The fraction of galaxies with an AGN above a threshold
luminosity increases substantially with SFR
  \citep[e.g.,][]{syme11sfx, syme13agnsf, raff11agnsf, june13agnsf},
and the average AGN accretion rate in galaxies scales with stellar
mass and redshift in a way that closely mirrors the star-forming
``main sequence'' \citep{mull12agnms}.  The average AGN luminosity in
interacting pairs of galaxies at low redshift increases with
decreasing separation, tracing precisely the behavior that is seen for
SF \citep{elli11agninter, elli13irpair}, and Seyfert galaxies show a strong correlation between
  AGN luminosity and {\em nuclear} SF  \citep[e.g.,][]{dima12agnsf, esqu13agnsf_aph}. Recently,
the connection between galaxy-scale SF and AGN activity was directly
tested by \citet{chen13agnsf}, who showed that when star forming
galaxies at $z\sim 0.5$ are divided up by SFR, their average BH
accretion rate is directly proportional to their SFR, consistent with
previous results for galaxies at higher redshift \citep{syme11sfx,
  raff11agnsf}.

These observations present the interesting puzzle that the
luminosities of individual AGNs show a weak or absent correlation with
the properties of their host galaxies, but that BH accretion appears
tightly linked to SF in a more global sense (for an extensive review
of these results see \citealt{alex12bh}).  In this paper, we explore a
possible solution, first suggested by \citet{alex12bh}, in which the
differences are caused by significant AGN variability on timescales
shorter than those characteristic of galaxy evolution (see also
  \citealt{neis13agnsf} for a detailed discussion of the
  observational effects of these different timescales). We construct
a simple model in which the growth rate of BHs (averaged over galaxy
evolution timescales of $\gtrsim$\,100 Myr) is exactly
proportional to the SFR in its host, but the instantaneous {\em
  observed} AGN luminosity can vary over a wide dynamic range, and
explore the observational consequences of this scenario. In
\S\,\ref{sec:variability} we discuss AGN variability, in
\S\,\ref{sec:model} we present the details of the model, in
\S\,\ref{sec:results} we present the implications of this model and
compare to observations, and in \S\,\ref{sec:dis} we discuss the
results in the context of the cosmological links between BHs and
galaxies.

\section{AGN variability}

\label{sec:variability}

\begin{figure}
\epsscale{1.15}
\plotone{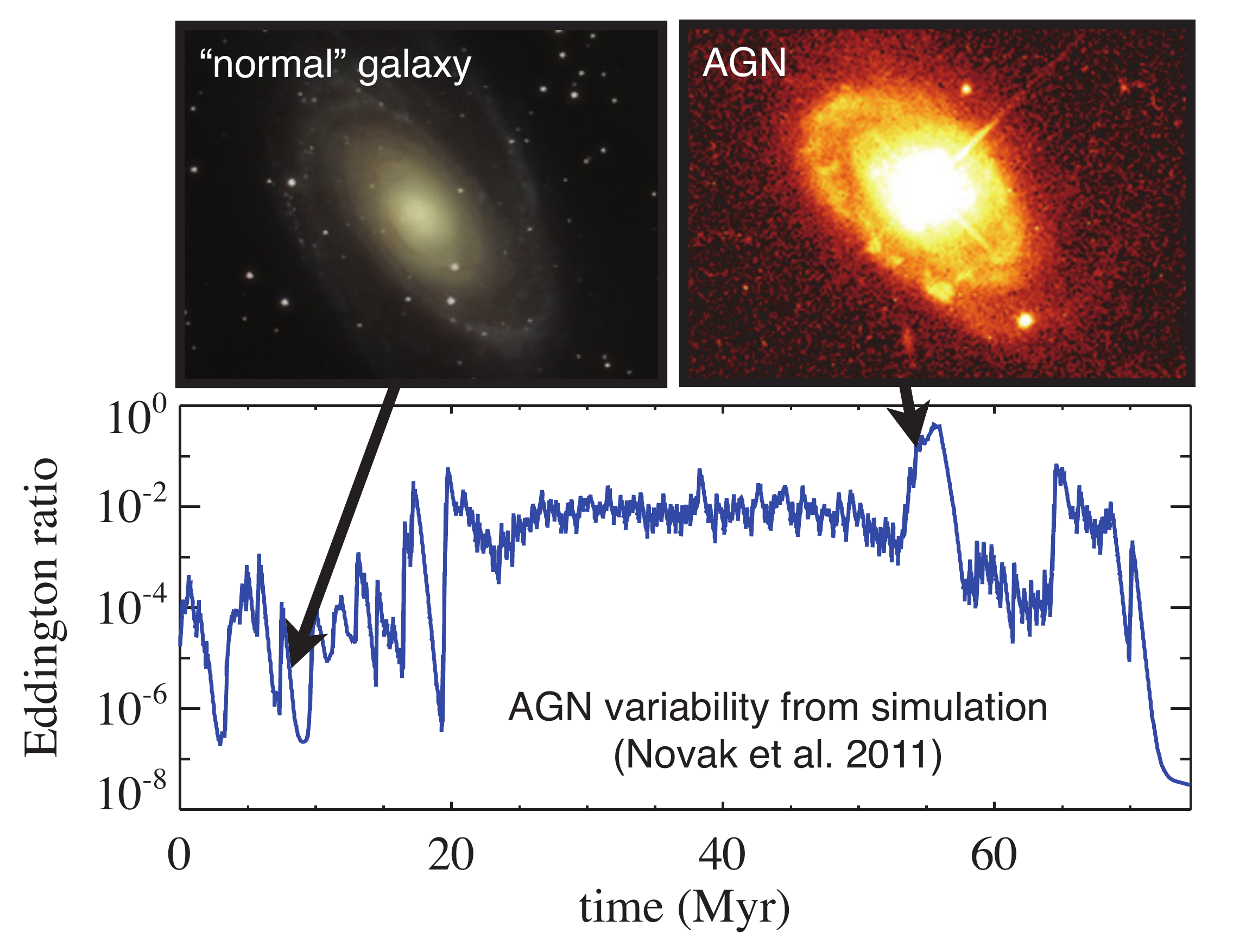}
\caption{Schematic illustration of AGN variability. The bottom panel shows the Eddington ratio as a function of time for one time interval in a hydrodynamic simulation presented by \citet{nova11bhsim}, shown in the bottom panel of their Figure 6. These simulations suggest that galaxies can switch between an ``inactive'' state (left panel) and a bright AGN state (right panel) over timescales of $
\sim$\,Myr or less. Image credits, from left: M81: Wikimedia Commons; PG 0052+251: J.\ Bahcall (IAS, Princeton), M.\ Disney (Univ.\ of Wales), NASA/ESA. 
\label{fig:schem}}
\end{figure}

The process of accretion onto BHs is known to be highly variable with
time. Stellar-mass BH binaries exhibit variations in the accretion
luminosity on a wide range of time scales, and are observed to change
dramatically in luminosity by $\gtrsim$\,5 orders of magnitude over
periods of days to weeks, owing to feedback from the accreting BH and
transitions in the mode of accretion \citep[e.g.,][]{chen97xrn,
  remi06bhb}. AGNs also show significant but smaller-amplitude (up to
2 orders of magnitude) variability on timescales from hours to years,
observed at optical/UV and X-ray wavelengths that probe
accretion disk and coronal emission, respectively \citep[e.g.,][]{ulri97agnvar,
  mcha12agnvar}.  However, this small-scale variability is {\em not}
analogous to accretion state changes in BH binaries, because the
dynamical and viscous timescales for accretion increase with BH mass,
such that an equivalent state transition that lasts several days in a
10 \msun\ binary might be expected to take $\sim$\,$10^4$ years for a
$\sim$\,$10^7$ \msun\ BH. It is natural to expect that AGNs, like
X-ray binaries, experience variability over a large dynamic range in
luminosity (Figure~\ref{fig:schem}), corresponding to changes in the
accretion state driven by various physical processes including
feedback on the accreting material \citep[e.g.,][]{hopk05life1,
  nova11bhsim} or accretion disk instabilities
\citep[e.g.,][]{siem97qsolf}. However, this variability could never be
directly observable for an individual AGN over the history of modern
astronomy.

There are, however, {\em indirect} measurements that give us clues to
longer-term AGN variability. Recently, AGN light echoes in the form of
large [\ion{O}{3}]-emitting clouds have been discovered on the
outskirts of several galaxies without clear evidence of ongoing AGN
activity \citep{lint09voor, scha10voor, keel12voor, keel12echo} or as
ultra-luminous galaxy-wide narrow-line regions
\citep{schi13echo}. These large ionized clouds have spectral
parameters indicating they were excited by AGN continuum radiation,
and the inferred AGN luminosity needed to illuminate these clouds
suggests that the source could have been many orders of magnitude (as
high as $10^5$ times) brighter at a time in the past corresponding to
the light travel time from the nucleus of $\sim$\,$10^4$ years
\citep{scha10voor, schi13echo}.  Similar evidence comes from the Milky
Way, where reflections of X-ray emission off molecular clouds indicate
that the Galactic Center varied in brightness by up to
$\sim$\,$10^{3}$ times over the past $\sim$\,500 years
\citep[e.g.,][]{pont10gcagn, cape12gcagn, gan13gcagn}, while the
observed ``{\it Fermi} bubbles'' observed in $\gamma$-ray emission may
be relics of an intense AGN phase $\sim$\,$10^6$ years ago
\citep{zubo11fermi, zubo12fermi, su12fermijet}. For luminous quasars
at high redshift, the transverse proximity effect provides a measure
of the variability based on the sizes of ionized bubbles around the
quasars, typically yielding lifetimes for luminous accretion of $\sim$\,$10^6$--$10^7$ years
\citep[e.g.][]{jako03prox, gonc08prox, kirk08prox}.  For AGNs with
  relativistic radio-bright jets, observations of hot spots and lobes
  in the jets, as well as cavities inflated by the mechanical
  outflows, suggest highly intermittent activity with timescales of
  $\sim$\,$10^4$--$10^7$ years \citep[e.g.,][]{mcna07araa,
    siem10outburst}. Together, these observations provide clear
  evidence that AGN accretion rates can vary by many orders of
  magnitude on timescales from hundreds to millions of years.

\begin{figure}
\epsscale{1.15}
\plotone{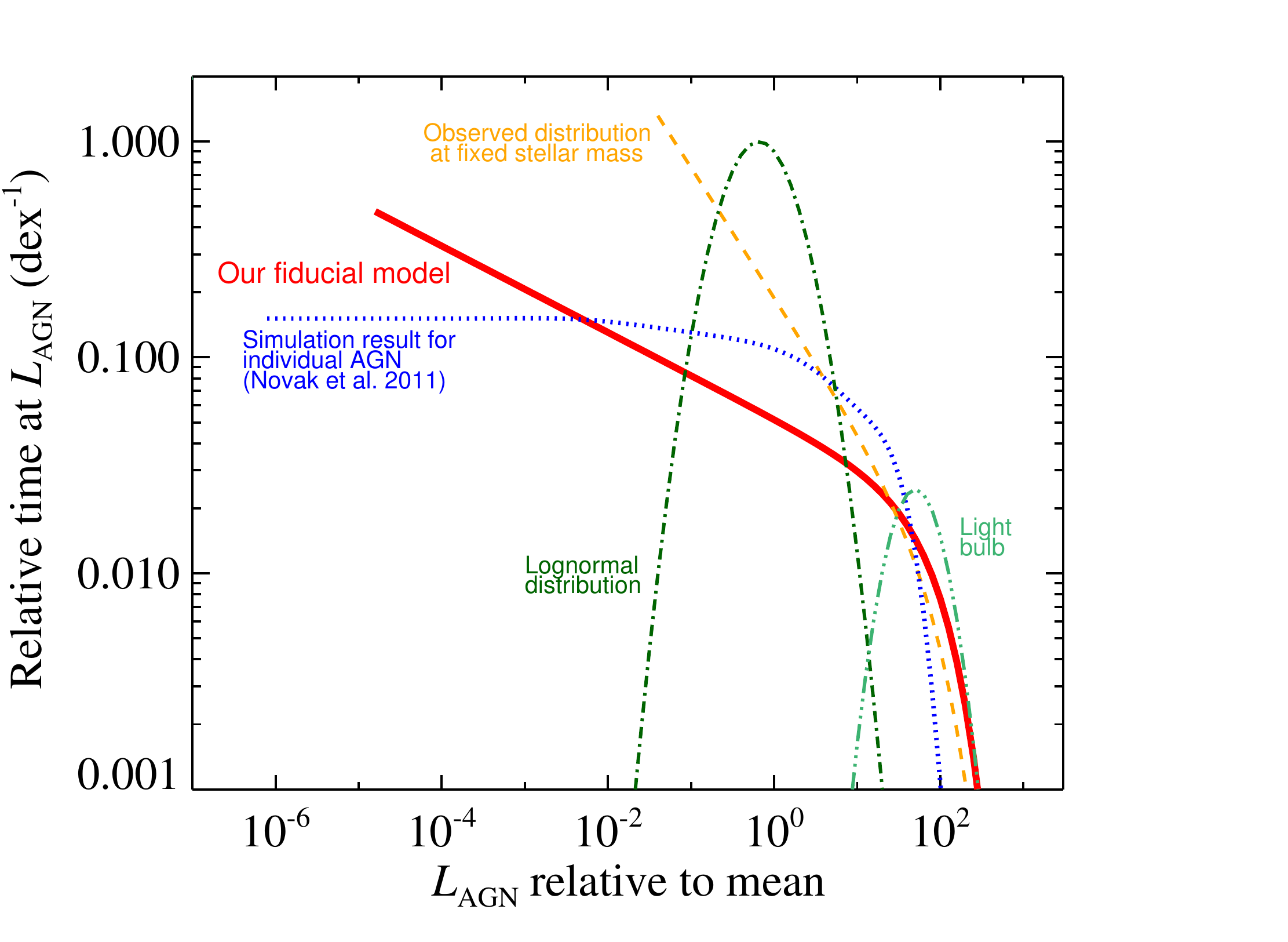}
\caption{Our fiducial model distribution of AGN luminosity compared to
  recent observational and theoretical results. The figure shows the
  relative amount of time spent per logarithmic interval in $L_{\rm
    AGN}$, plotted in terms of its value relative to the average
  long-term AGN luminosity. The yellow dashed line shows the Schechter
  function distribution \citep[Equation~(\ref{eqn:plaw})]{hopk09bulb}
  with power law index $\alpha=0.6$ that approximately matches
  observed Eddington ratio and specific accretion rate distributions
  \citep{hopk09bulb, aird12agn, bong12xgal}, and consistent with optical
spectroscopic studies of AGN in low-redshift passive galaxies
\citep{kauf09modes}. The green dot-dashed line shows a
lognormal distribution, as obtained by \citet{kauf09modes} for AGNs in
low-redshift star-forming galaxies, while the light green
dot-dot-dashed line shows a ``light bulb'' model with a duty cycle
$\sim$\,$10^{-2}$ \citep[e.g.,][]{conr13qso}. The blue dotted line
shows the distribution from the hydrodynamic simulations of
\citet{nova11bhsim}, which can be approximately modeled by an
exponential (i.e., a Schechter function with $\alpha=0$).  The
fiducial model we adopt in this paper consists of Schechter function
with $\alpha=0.2$; this model gives a qualitatively good match to
observations of the SF and AGN activity as discussed
below. Details of the different distributions are presented in
\S\,\ref{sec:variability}. We discuss the observational consequences
of changing the adopted luminosity distribution in
\S\,\ref{sec:distribution}.
\label{fig:bhar}}
\end{figure}

A large dynamic range of variability is consistent with the wide
distribution of Eddington ratios observed for AGNs in extragalactic
surveys. Studies of nearby optically-selected AGNs found that
star-forming and passive galaxies have characteristic log-normal and
power-law distributions of Eddington ratio that were
independent of BH mass \citep{kauf09modes}. \citet{hopk09bulb} derived Eddington ratio distributions
using a range of AGN observations and showed that they can be modeled
with a theoretically-motivated functional form consisting of a
Schechter function, defined as a power law with an exponential cutoff
near the Eddington limit:
\begin{equation}
\frac{{\rm d}t}{{\rm d}\log L} = t_0 \left( \frac{L}{L_{\rm cut}}\right)^{-\alpha} \exp(-L/L_{\rm cut}).
\label{eqn:plaw}
\end{equation}

\citet{hopk09bulb} found that the observed Eddington ratios can be
reproduced with $L_{\rm cut} \approx 0.4 L_{\rm Edd}$ and
$\alpha\approx 0.6$. This power-law slope is similar to that found by
\citet{kauf09modes} for optically-selected AGNs in low-redshift
passive galaxies. We note that recent studies suggest that many
  low-ionization sources included in the low-Eddington optical AGN
  population are likely to be powered by evolved stellar
  populations, rather than BH accretion \citep[e.g.,][]{sarz10liner,
    yan12liner, sing13liner}, although the effects of this
  contamination on the derived Eddington ratio distribution are
  difficult to assess. However, X-ray selected AGNs at moderate to
  high redshift (which suffer little contamination from sources
  powered by evolved stars) show a similar universal power-law shape
  for the specific accretion rate distribution, parameterized by the
  ratio of $L_{\rm AGN}$ to stellar mass. This distribution is
independent of stellar mass, but increases in amplitude with redshift
similarly to the evolution of the star-forming ``main sequence'',
suggesting a connection between AGN accretion and SF \citep{aird12agn,
  bong12xgal}. \citet{aird12agn} obtained a power-law slope to this
distribution of $\alpha \approx 0.6$, similar to that measured by
\citet{hopk09bulb}. \citet{bong12xgal} report a somewhat steeper slope
($\alpha \approx 1$), although this is measured over a smaller dynamic
range in specific accretion rate.  For luminous broad-line quasars,
recent studies have directly measured Eddington ratios using the
virial technique.  Correcting for incompleteness effects, these
analyses have yielded similarly broad distributions that cut off at
high Eddington ratios and increase to low accretion rates, although
the slopes of these distributions on the low end are less
well-constrained than for X-ray selected samples
\citep[e.g.,][]{nobu12edd, kell10qsoedd, kell13edd}.

For the purposes of the discussion below, we will represent the
``observed'' AGN luminosity distribution at fixed stellar (or BH) mass
using the \citet{hopk09bulb} model given in Equation (\ref{eqn:plaw})
with $\alpha=0.6$ and $L_{\rm cut} = 100 \average{L_{\rm AGN}}$. This
luminosity distribution is shown as the dashed orange line in
Figure~\ref{fig:bhar}, where $L_{\rm AGN}$ is plotted in terms of its
value relative to the long-term average luminosity $\average{L_{\rm
    AGN}}$. We normalize this distribution (by scaling $t_0$) such
that the integral over all $L_{\rm AGN}$ is equal to 1. The curve
therefore represents the {\em relative} time spent by the AGN in each
logarithmic interval of luminosity. This value of $L_{\rm cut}$
adopted here, while somewhat arbitrary, is motivated by two main
considerations. First, it reproduces well the observed relationship
between SFR and BH accretion rate in powerful AGNs, as shown in
\S\,\ref{sec:results}. Second, for a Milky Way-type galaxy and our
adopted model parameters (as we discuss in \S\,\ref{sec:model}), this
value of $L_{\rm cut} \approx 0.4 L_{\rm Edd}$, consistent with the
cutoff adopted by \citet{hopk09bulb} to fit observed Eddington ratio
distributions.  We note that, by definition, the average luminosity of
the distribution equals $\average{L_{\rm AGN}}$, which requires that
the distribution be truncated at some lower limit. For $L_{\rm cut} =
100 \average{L_{\rm AGN}}$ and $\alpha=0.6$, we require a lower limit
at $L_{\rm AGN} \approx 10^{-2} \average{L_{\rm AGN}}$, as shown in
Figure~\ref{fig:bhar}. This distribution therefore represents
variability of the AGN accretion rate over approximately four orders
of magnitude, consistent with the lower limit on the dynamic range
observed by a number of studies \citep[e.g.][]{hopk09bulb,
  kauf09modes, aird12agn, bong12xgal}.

In addition to broad power-law or Schechter function models, several studies
have considered narrower distributions for the AGN accretion rate.
\citet{kauf09modes} found that optically-selected AGNs in galaxies
with young stellar populations show an approximately lognormal
distribution in accretion rate (with width $\sigma \approx 0.4$ dex)
that is independent of mass.  The lognormal distribution is given by
\begin{equation}
\frac{{\rm d}t}{{\rm d}\log L} = t_0 \exp \left(- \frac{(\log L - \log L_0)^2}{2 \sigma^2} \right)
\label{eqn:gauss}
\end{equation}
and is shown as the dot-dashed dark green line in
Figure~\ref{fig:bhar}. We again scale $t_0$ so that this curve
represents the relative time spent in each dex of $L_{\rm AGN}$. The
centroid luminosity $L_0$ depends on $\sigma$, to ensure that the
average accretion rate in the distribution is equal to
$\average{L_{\rm AGN}}$. For a width $\sigma = 0.4$ dex, $L_0 \approx
0.8 \average{L_{\rm AGN}}$, as shown in Figure~\ref{fig:bhar}.

A number of works have also considered a ``light bulb'' scenario in
which every galaxy spends a fixed fraction of time ``on'' as an AGN
(with some distribution of accretion rates) and the remainder as a
``normal'' galaxy at very low Eddington ratio.  Employed in a
cosmological context, such models can be effective at modeling the
quasar luminosity function (LF), clustering, and other observables
\citep[e.g.][]{shan13agn, conr13qso}, although they do not reproduce
observed Eddington ratio distributions for moderate-luminosity AGNs
\citep[e.g.,][]{hopk09bulb, shan13agn}. The recent model of
\citet{conr13qso} adopted a lognormal Eddington ratio distribution
with $\sigma = 0.3$ dex, for which fits to the quasar LF yielded a duty
cycle (the fraction of time the quasar is ``on'') of
$\sim$\,$10^{-3}$--$10^{-2}$ at $z < 3$. In this study we model such a
``light bulb'' scenario using a lognormal distribution with
$\sigma=0.3$ and $L_0 \approx 50 \average{L_{\rm AGN}}$ (chosen to
approximately match the SFRs of luminous AGNs, as shown in
\S\,\ref{sec:distribution}), with a corresponding duty cycle
$\sim$\,$10^{-2}$. In the model, objects in the ``off'' state are set
to an arbitrarily low $L_{\rm AGN} \ll \average{L_{\rm AGN}}$, the
precise value of which does not affect the results. This ``light
bulb'' distribution is shown as the light green dot-dot-dashed line in
Figure~\ref{fig:bhar}.

\begin{figure*}
\epsscale{1.15}
\plottwo{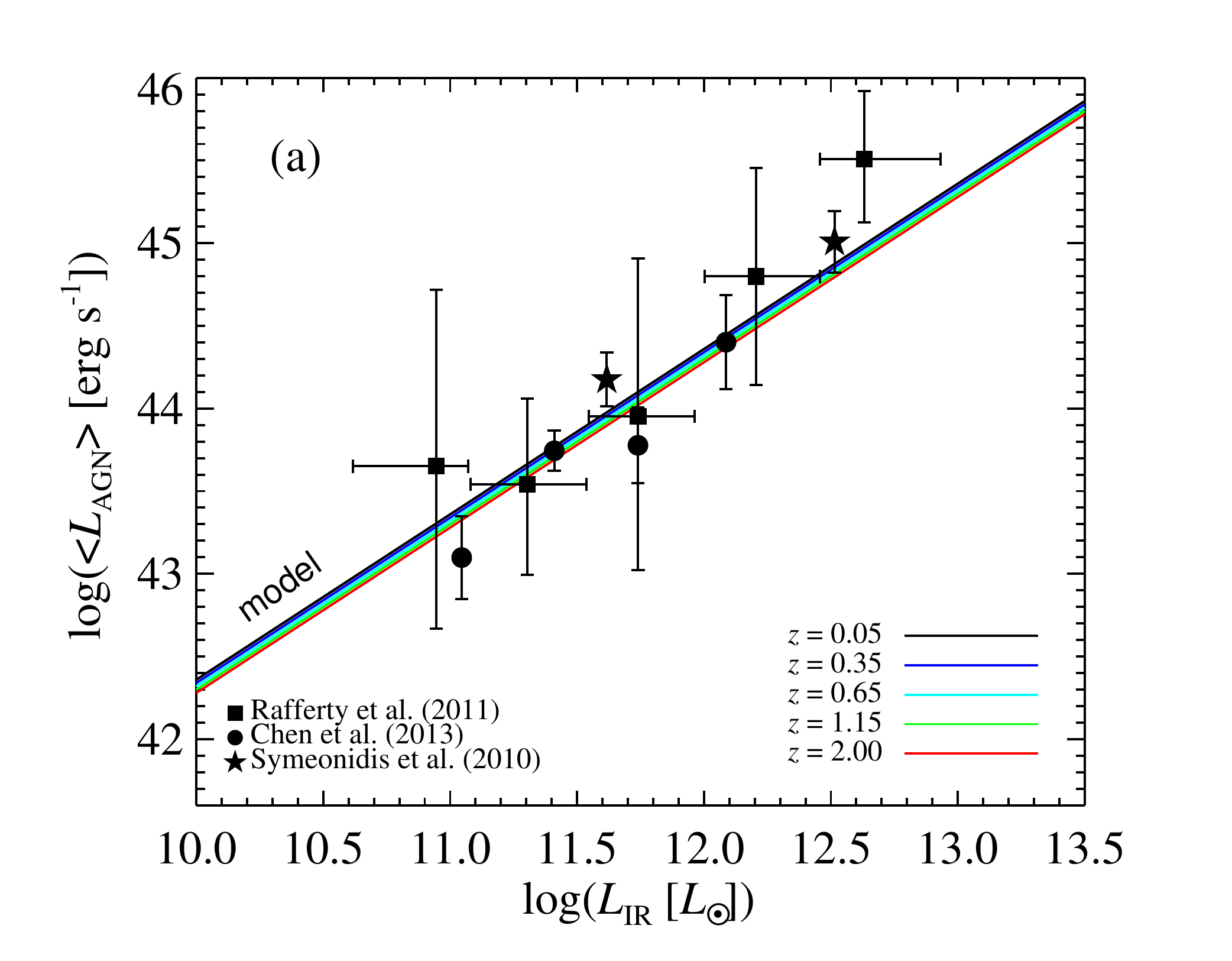}{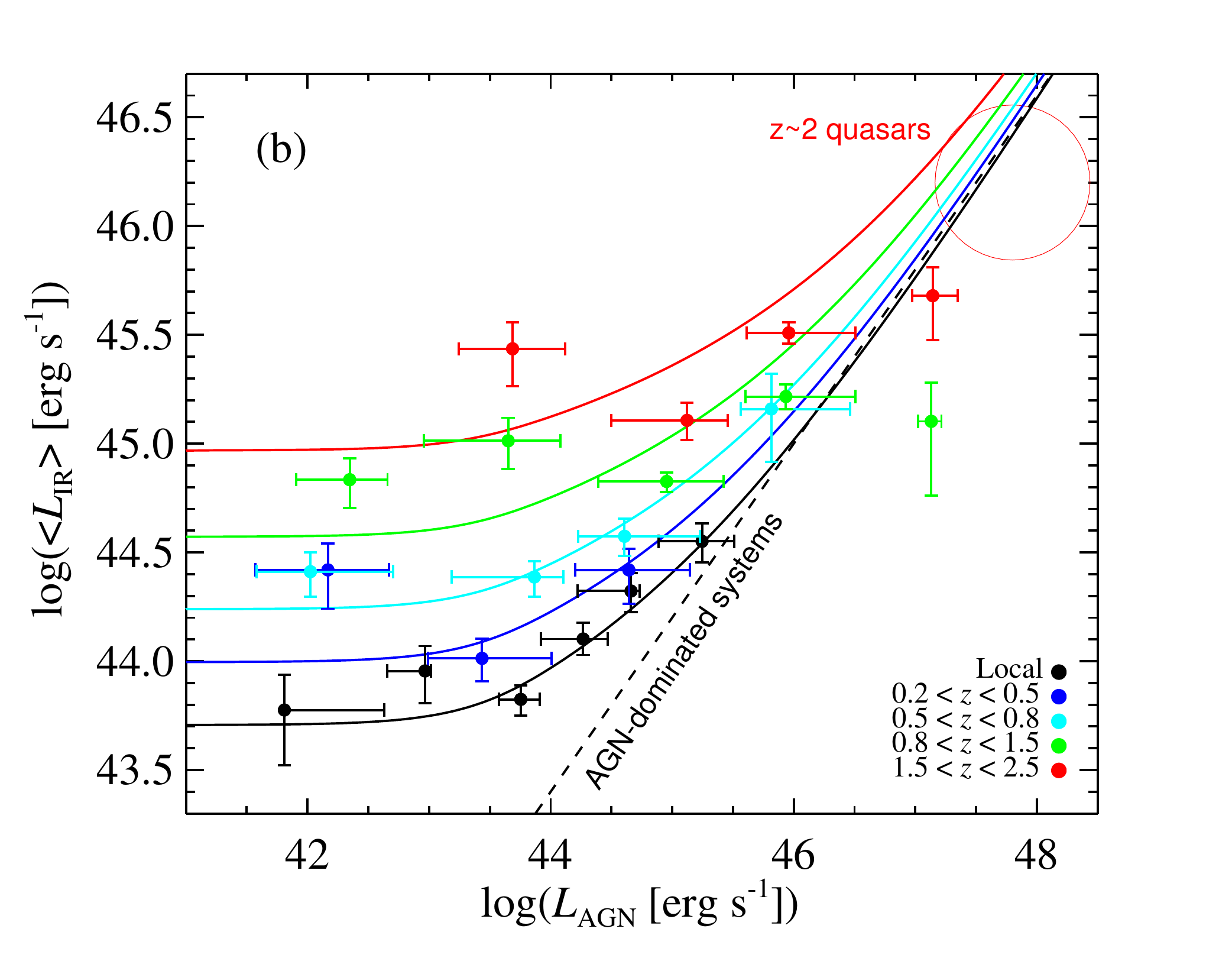}
\caption{Comparison of the predictions of our AGN variability model to
  observations of SF and AGN activity in galaxies. (a) The average AGN
  luminosity for star-forming galaxies as a function of the observed
  \lir. The model predictions for each redshift are shown by the
  colored lines. By design, these show a constant one-to-one
  correlation between $\average{L_{\rm AGN}}$ and \lir\ at all
  redshifts (the lines are offset for clarity), independent of the
  choice of accretion rate distribution. The points show recent
  observational results, which are consistent with this simple
  picture. (b) The average \lir\ for AGN host galaxies, as a 
  function of \lagn. Data points are taken from \citet{rosa12agnsf},
  and solid lines show the predictions of our fiducial model. The
  models curves are evaluated at the central redshift of each range;
  the ``local'' model which is calculated for $z=0.05$, the median
  redshift for the {\em Swift}-BAT AGNs used in the analysis
  \citep{rosa12agnsf, cusu10bat}.  The red circle shows the approximate range of
  \lir\ observed for quasars at $z\sim2$ \citep{lutz08qsosf} and the
  dashed line shows the approximate relation between AGN and IR
  luminosity for AGN-dominated systems ($L_{\rm IR} \propto L_{\rm
    AGN}^{0.8}$) determined by \citet{netz09agnsf}. Our simple model
  clearly reproduces the general trends in the \citet{rosa12agnsf}
  data while also matching the close relation between $L_{\rm AGN}$
  and \lir\ for AGN-dominated systems \citep{lutz08qsosf,
    netz09agnsf}.
\label{fig:agnvsf}}
\end{figure*}

Several studies have also considered accretion rate distributions that
are a sum of power-law and lognormal components. Such a distribution
was obtained for the full sample of galaxies (passive and
star-forming) studied by \citet{kauf09modes} using optical
spectroscopy, and was favored by \citet{shan13agn} in order to best
reproduce Eddington ratio distributions, AGN LF, and other properties
of the AGN population across a range of redshifts. The observable quantities (as discussed in
\S\,\ref{sec:results}) that are predicted by such combined models are
similar to those produced by a pure Schechter function, although the
details depend on the relative normalizations of the lognormal and
power-law components. For simplicity we restrict this paper to
Schechter and lognormal distributions and do not consider such
``combined'' models in further detail.

In addition to the observational results discussed above, theoretical
understanding of the luminosity distribution of AGNs may be obtained
from recent simulations of BH accretion including small-scale feedback
\citep[e.g.,][]{hopk05life1, ciot10flare, nova11bhsim, gabo13bhsim} or a
detailed treament of gravitational torques
\citep[e.g.,][]{angl13bhsim,angl13bhsim2_aph} that show strong
stochastic variations in the BH accretion rate over a range of
timescales. The accretion rate distribution from the simulation of
\citet{nova11bhsim} is shown in Figure~\ref{fig:bhar}. These
simulations shows that the accretion rate for a given BH can vary by
more than 7 orders of magnitude in a timescale of $\sim$\,Myr or
less. The model BH spends approximately equal time in any given
logarithmic interval of accretion rate, with a cutoff at high
accretion rates due to the Eddington limit. This distribution can be
approximately described by an exponential (that is, a Schechter
function as for the observational results discussed above, but with
$\alpha = 0$). Similar results were obtained in recent simulations by
\citet{gabo13bhsim}, who found that the characteristic power law slope
of the accretion rate distribution varies depending on gas fraction,
from $\alpha \approx 0.6$ for a galaxy like the Milky Way to
$\alpha \approx 0.2$ for systems with a higher gas fraction
characteristic of galaxies at $z=2$.

On the whole, Figure~\ref{fig:bhar} illustrates the wide diversity in
AGN accretion rate and luminosity distributions that are obtained from
theoretical models and observational studies of different galaxy
populations.  For most of what follows we will adopt a fiducial
distribution that lies between the \citet{nova11bhsim} theoretical and
\citet{hopk09bulb} observational curves, following the Schechter
functional form in Equation (\ref{eqn:plaw}), with $\alpha = 0.2$ and
$L_{\rm cut} = 100 \average{L_{\rm AGN}}$.  As shown in
Figure~\ref{fig:bhar}, this distribution requires a lower limit to
$L_{\rm AGN}$ of $\sim10^{-5} \average{L_{\rm AGN}}$.  We note that
the precise value of $L_{\rm cut}$ makes no significant difference to
our results; increasing or decreasing $L_{\rm cut}$ by a factor of two
changes the model predictions by less than the observational
uncertainties on the various quantities discussed in
\S\,\ref{sec:results}.

As we demonstrate in \S\,\ref{sec:results}, our fiducial distribution
displays a good qualitative agreement with a range of observational
data. However, we stress that in this study we do {\em not} attempt to
constrain the precise distribution in AGN luminosity or estimate
uncertainties. Such an analysis would require a very careful
understanding of the biases and uncertainties in the various
observational constraints and will likely suffer from significant
degeneracies in the functional form for the distribution in $L_{\rm
  AGN}$. Rather, the scope of this work is limited to showing that
with a reasonable choice of the luminosity distribution, short-term
variability of the AGN can explain a range of observations even if BH
accretion and SF are perfectly correlated on timescales typical of
galaxy evolution. We describe this simple model in the next section.

\section{A simple model connecting black hole accretion and star formation}

\label{sec:model}

Motivated by our knowledge of AGN variability and the observed and
theoretical links between SF and BH accretion, we may ask
a straightforward question: Observed over timescales typical of star
formation ($\gtrsim$\,100 Myr; e.g., \citealt{hick12smg}) do {\em all} star-forming galaxies host an
AGN?  Here we construct a simple model in which SF and
(long term) BH growth are perfectly correlated in galaxies,
and explore the observational consequences. The ingredients of the
model are as follows:

\begin{enumerate}

\im We create a model population of star-forming galaxies across a
range of redshifts from 0 to 2, with a redshift-dependent distribution
in SFR taken from the far-IR LF derived by \citet{grup13irlf} from
{\em Herschel} observations. We convert far-IR luminosity to SFR using
the prescription of \citet{kenn98araa}, multiplying by a factor of 0.6
as appropriate for a \citet{chab03imf} initial mass function (IMF).
We include galaxies with total far-IR luminosities $10^9 < L_{\rm
    IR} < 10^{14}$ $L_\odot$, covering the full luminosity range that
  is well-constrained by the {\em Herschel} LF measurements at any
  redshift \citep{grup13irlf} and corresponding to SFRs characteristic
  of essentially all star-forming galaxies ($\sim$0.1--$10^4$
  $M_\odot$ yr$^{-1}$). Varying these limits in $L_{\rm IR}$ by an
  order of magnitude has no effect on our conclusions.

\im For each galaxy, we assign an {\em average} BH accretion rate
(i.e., averaged over $\gtrsim$\,100 Myr) such that ${\rm SFR}/{\rm
  BHAR} = 3000$, motivated by the observed ratios of SF
and BH accretion rates \citep{raff11agnsf, mull12agnms, chen13agnsf}. 

\im For each galaxy, we assume that the {\em instantaneous} accretion
rate relative to the average is drawn from the fiducial luminosity
distribution shown in Figure~\ref{fig:bhar}. We note that the adopted
distribution is in {\em luminosities} rather than Eddington ratios,
because for simplicity we do not account  for the BH mass in
each model galaxy.  We also stress that the particular timescales over which
these variations occur (and thus the precise variability power
spectrum) are not important in this model. We require only that the
fluctuations occur on timescales significantly shorter than a typical
SF timescale of $\gtrsim$\,100 Myr, so that the distribution of AGN
luminosities is well-sampled at any given SFR.

\im We convert the instantaneous accretion rate $\dot{m}_{\rm BH}$ to
a bolometric AGN luminosity via $L_{\rm AGN} = \epsilon {\dot{m}_{\rm BH}} c^2$,
assuming a constant radiative efficiency $\epsilon = 0.1$.\footnotemark

\setcounter{footnote}{3}

\footnotetext{Assuming $\epsilon = 0.1$ and ${\rm SFR}/{\rm BHAR} =
  3000$, a Milky Way-type galaxy with ${\rm SFR} = 1\, M_\odot$
  yr$^{-1}$ \citep[e.g.,][]{roba10mwsfr} will have $\average{L_{\rm
      AGN}} \approx 2\times10^{42}$ \ergs. The Eddington limit for a
  $4\times10^6$ \msun\ BH \citep{ghez08sgra, gill09sgra} is $L_{\rm
    Edd} \approx 5\times10^{44}$ \ergs, so that for $L_{\rm cut} = 100
  \average{L_{\rm AGN}}$, $L_{\rm cut} \approx 0.4 L_{\rm Edd}$ for
  the Milky Way as discussed in \S\,\ref{sec:variability}.}

\end{enumerate}

This procedure yields a model population of AGNs and galaxies with an
an instantaneous BH accretion rate (given by $L_{\rm AGN}$) and SFR
(given by $L_{\rm IR}$) that can be compared to observations. 
With this simulated population of galaxies and AGNs in hand, we can
then calculate the same relationships between SF and AGN luminosity
that have been derived from recent extragalactic surveys. 

\section{Results}
\label{sec:results}

\subsection{Star formation rate and $L_{\rm AGN}$}
\label{sec:sfagn}
We begin by calculating the average AGN luminosity for galaxies in
bins of $L_{\rm IR}$ for a range of redshifts from 0 to 2. This
process serves to average over the variability of the AGNs for a given
SFR, and so by the design of the model, produces an exactly
proportional relationship between $L_{\rm IR}$ and $\average{L_{\rm
    AGN}}$ that is independent of redshift, as shown in
Figure~\ref{fig:agnvsf}(a). This relationship agrees with that
observed in the recent studies of the average BH accretion rate in
star-forming galaxies \citep{syme11sfx, raff11agnsf,
  chen13agnsf}. For the \citet{syme11sfx} result we have computed
the $\average{L_{\rm AGN}}$ in bins of $L_{\rm IR}$ from the data
points reported in their paper, as discussed in
\citet{chen13agnsf}.

We next perform the opposite calculation, and compute the average
\lir\ as a function of \lagn. In the context of our model, this
entails selecting galaxies based on the {\em unstable}, rapidly
varying quantity (BH accretion rate or $L_{\rm AGN}$) and averaging over
the {\em stable} quantity (SFR or $L_{\rm IR}$). This analysis is
motivated by several recent measurements that have found weak
correlations between average SF luminosity and $L_{\rm AGN}$
\citep{lutz10agnsf, shao10agnsf, rosa12agnsf, harr12agnsf}.  The
observed data points from \citet{rosa12agnsf} are shown in
Figure~\ref{fig:agnvsf}(b), converting from $\nu L_\nu$ at 60 microns
(as presented in their work) to \lir\ (integrated from 8---1000
$\mu$m) by adding 0.2 dex, as is typical for the average spectral energy distributions of
star-forming galaxies \citep{char01sfgal, kirk13agnsf}. We also show
the observed relationships for optically-selected AGN-dominated systems
\citep{netz09agnsf} and for high-redshift quasars \citep{lutz08qsosf}.

\begin{figure}
\epsscale{1.15}
\plotone{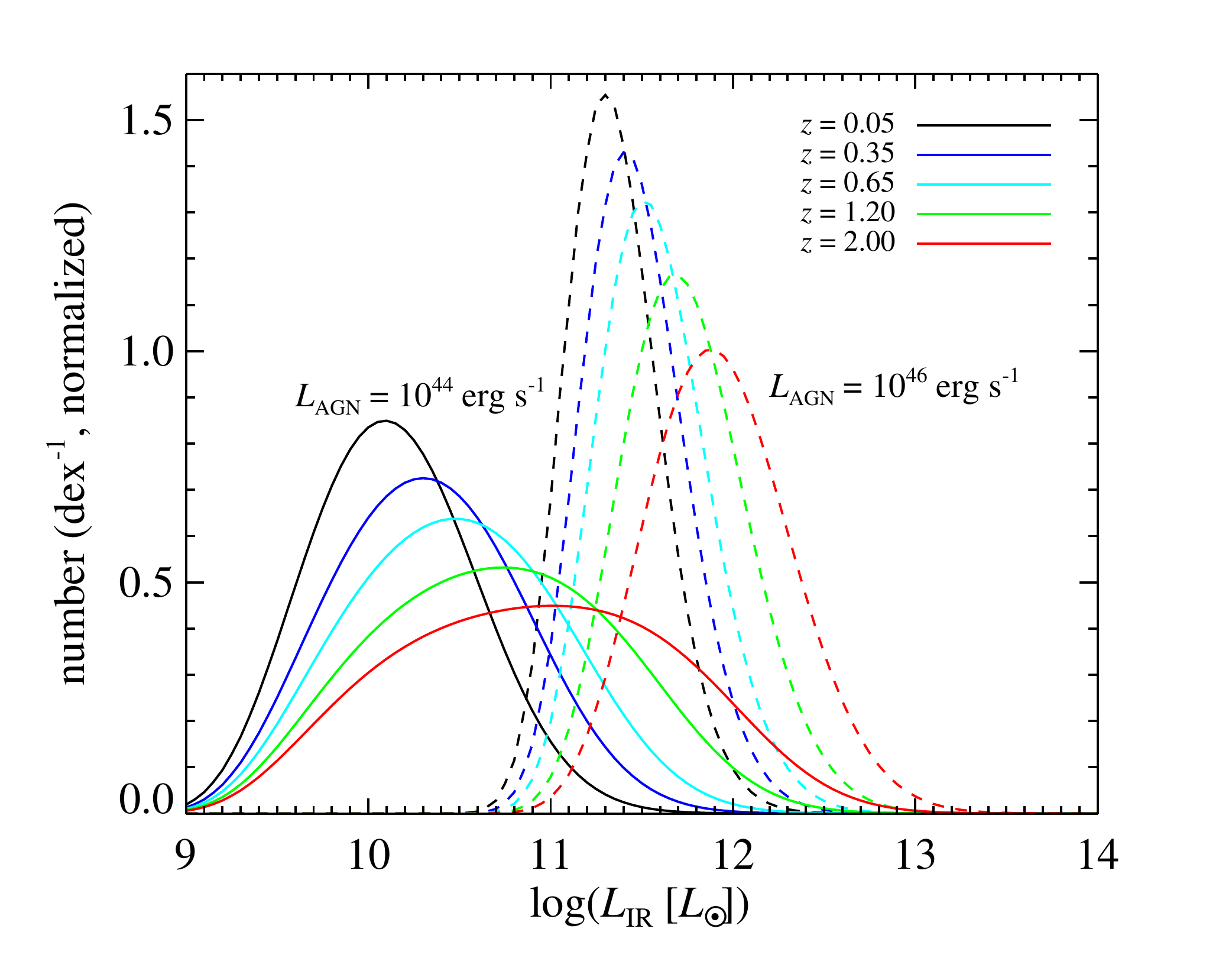}
\caption{Distributions of $L_{\rm IR}$ predicted by our simple model
  for AGNs  with moderate $L_{\rm AGN} = 10^{44}$ \ergs\ (solid lines)
  and high $L_{\rm AGN} = 10^{46}$ \ergs\ (dashed lines),
  evaluated at the same redshifts (coded by color) as in
  Figure~\ref{fig:agnvsf}. The distributions are normalized so that
  the integral under each curve is equal to 1.  This figure
  demonstrates two key aspects of our model for the connection between
  SF and AGN activity: a shift to higher $L_{\rm IR}$ with increasing
  redshift at fixed $L_{\rm AGN}$, and the narrow distribution of
  $L_{\rm IR}$ for high-luminosity AGNs compared to systems with lower
  $L_{\rm AGN}$. \label{fig:sfdist}}
\end{figure}

To compare our simple model to these results, we select our simulated
AGNs in bins of their observed (instantaneous) $L_{\rm AGN}$, and then
average the $L_{\rm IR}$ for the objects in each bin. The model
results are shown as solid lines in Figure~\ref{fig:agnvsf}(b), and
closely match the key features of the observations. In particular,
there is no strong correlation between $\average{L_{\rm IR}}$ and
$L_{\rm AGN}$ for moderate-luminosity AGN, with each luminosity having
an average $L_{\rm IR}$ corresponding to the typical SFR (the ``knee''
of the IR LF) at each redshift. This average correspondingly shifts to
higher luminosity with redshift as the typical SFR of galaxies
increases \citep[e.g.,][]{noes07, elba11ms}. 
  Figure~\ref{fig:sfdist} shows the predicted distribution of $L_{\rm
    IR}$ for various $L_{\rm AGN}$ and redshift,  demonstrating
  this shift to higher $L_{\rm IR}$ for increasing $z$.

Another important feature of the model is the emergence of a
correlation between $\average{L_{\rm IR}}$ and $L_{\rm AGN}$ at high
luminosity. In the model, the highest AGN luminosities can only be
produced by galaxies with high SFRs that are near the high end of the
accretion rate distribution, so that high-luminosity AGNs are
  associated with narrower distributions in $L_{\rm IR}$ than lower-luminosity
  AGNs (as shown in Figure~\ref{fig:sfdist}).  Thus there emerges a
strong correlation between $\average{L_{\rm IR}}$ and $L_{\rm AGN}$,
as observed in a number of studies \citep[e.g.,][]{lutz08qsosf,
  netz09agnsf, bonf11qsosf}. We note that this correlation is not due
to contamination of the IR emission by the nucleus, as the AGN
contribution at far-IR wavelengths is minimal \citep{netz07qsosf,
  mull11agnsed, rosa12agnsf}.  In the model, the luminosity at which
we begin to see a strong correlation between $L_{\rm IR}$ and $L_{\rm
  AGN}$ increases with redshift as the ``knee'' of the IR LF shifts to
higher luminosities, explaining the fact that the weak correlations
between $\average{L_{\rm IR}}$ and $L_{\rm AGN}$ extend to higher
$L_{\rm AGN}$ in the observations at higher $z$.

Overall, this simple model produces remarkable agreement with both the
results on the average AGN luminosity of SF galaxies
\citep{syme11sfx, raff11agnsf, chen13agnsf} and the average
SFR of AGNs \citep{lutz10agnsf, shao10agnsf, rosa12agnsf}. We conclude
that the current observations are consistent with a picture in which
SF and BH accretion are closely connected over long timescales, but
this correlation is hidden at low to moderate $L_{\rm AGN}$ due to the short-term AGN variability.

\subsection{The AGN luminosity function}
\label{sec:lf}

\begin{figure*}
\epsscale{0.8}
\plotone{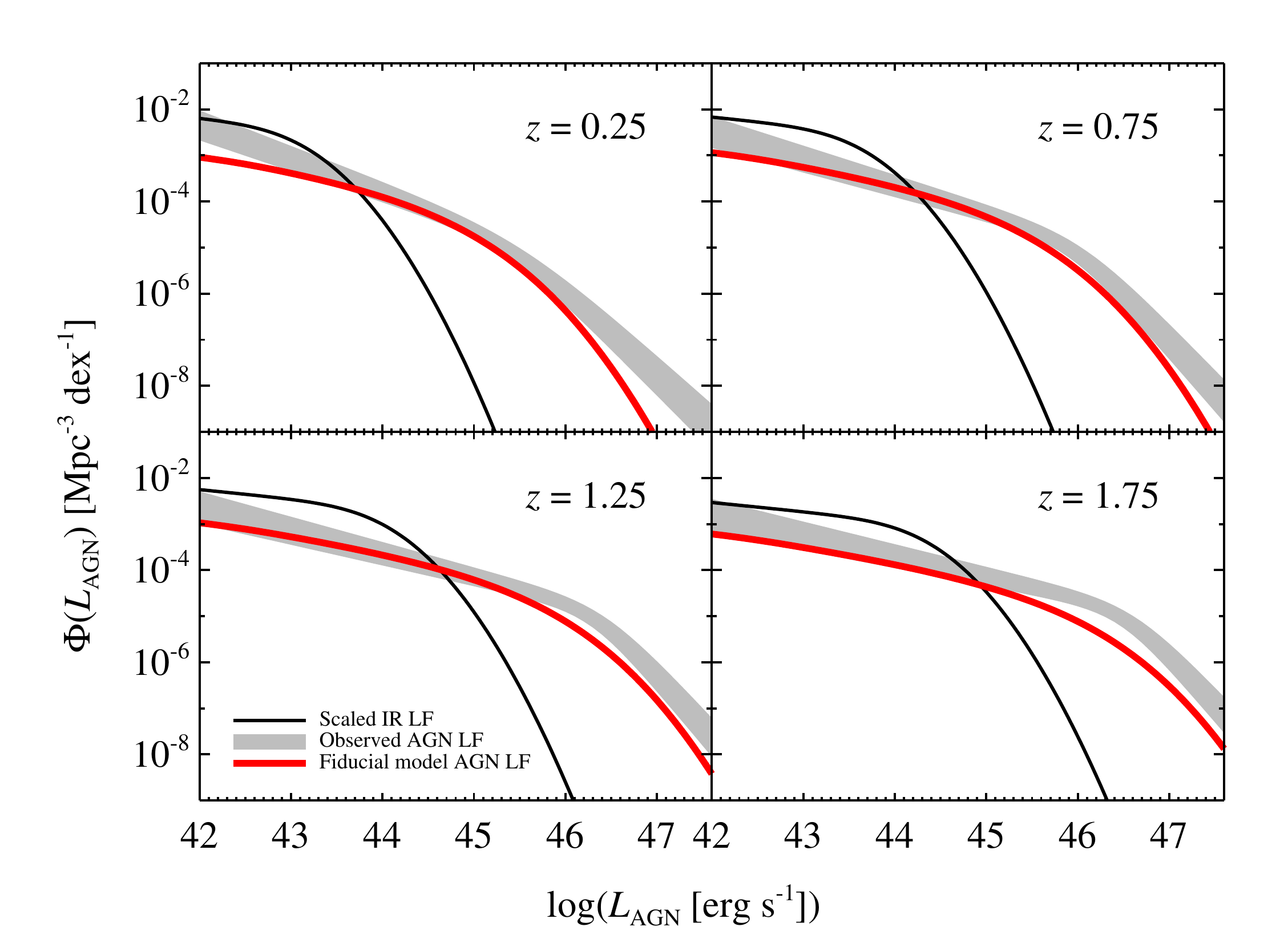}
\caption{Predictions of our simple model for the AGN luminosity
  function. Black lines show the distribution in {\em average} AGN
  luminosity, determined by scaling the observed IR LF
  \citep{grup13irlf} to the corresponding $L_{\rm AGN}$ as described
  in \S\,\ref{sec:lf}. The red lines show the predicted AGN LF from our
  fiducial variability model, derived by convolving the average AGN LF
  with the luminosity distribution shown in Figure~\ref{fig:bhar}. The
  shape and evolution of the LF predicted by the model agrees
   well with the observed AGN bolometric LF determined by
  \citet{hopk07qlf}, as shown by the gray shaded region. We use the
  fitting formula from \citet{hopk07qlf} for the ``full'' redshift
  evolution of $\phi(L_{\rm AGN})$, and estimate the uncertainties on
  the LF parameters based on typical uncertainties in fits to
  individual redshift ranges, given in Table 2  of \citet{hopk07qlf}.
  This result clearly shows that the differences between the
  characteristic shapes of the IR and AGN LFs can be largely explained by AGN
  variability. \label{fig:lf}}
\end{figure*}

As a further check on our simple prescription, we determine the LF of
AGNs in our model and compare to observations. A number of previous
studies have used the observed AGN LF to place constraints on the BH
accretion rate distribution \citep[e.g.,][]{siem97qsolf, hopk05life1,
  merl08agnsynth, shen12qsodem, conr13qso}. In our model, we produce a
predicted AGN LF by scaling the observed IR LF from \citet{grup13irlf}
to the expected AGN luminosities assuming no variability, applying the
appropriate factors to convert from $L_{\rm IR}$ to SFR and then to
$L_{\rm AGN}$, via
\begin{equation}
\average{L_{\rm AGN}} = \left(\frac{L_{\rm IR}}{C_{\rm IR}} \right) \left(\frac{1}{3000} \right)  \epsilon c^2,
\end{equation}
where $C_{\rm IR}$ is the $L_{\rm IR}$ to SFR conversion factor from
\citet{kenn98araa} assuming a Chabrier IMF, and the radiative
efficiency $\epsilon = 0.1$ as discussed above. We then convolve the
distribution in $\average{L_{\rm AGN}}$ with our adopted distribution
in accretion rates, to obtain the predicted ``observed'' distribution
in $L_{\rm AGN}$. We compare to the bolometric AGN LF determined by
\citet{hopk07qlf}. 

The observed AGN LF along with the model distributions in
$\average{L_{\rm AGN}}$ and ``observed'' $L_{\rm AGN}$ are shown in
Figure~\ref{fig:lf}, for four different redshifts. Approximate
uncertainties on $\phi(L_{\rm AGN})$ from \citet{hopk07qlf} are shown
by the gray shaded region. The model reproduces the general shape and
redshift evolution of the AGN LF remarkably well, particularly around
the ``knee'' where the LF is best constrained.  We note that at all
redshifts the model slightly underpredicts the observed LF, but this
may be expected; as we discuss in \S\,\ref{sec:dis}, clustering and
population studies indicate a non-negligible fraction of AGNs residing
in passive galaxies that would not be accounted for in a model that
ties all accretion to SF.

Nonetheless, our model may explain why the AGN LF shows a broader
high-luminosity tail compared to IR LF. In the model, each galaxy with
a high SFR can have a fairly wide range of $L_{\rm AGN}$, which serves
to flatten the luminosity distribution and extend it to higher
luminosities. The observed AGN LF is thus broadly consistent with the
view that AGN activity is ubiquitous in star-forming galaxies and
directly follows SF over long timescales, but with
significant stochastic variability.

\subsection{Merger fractions}

Another interesting test of our simple picture is a comparison to the
fraction of AGNs observed in galaxy mergers. A number of
theoretical models predict that AGN fueling is primarily driven by
mergers \citep[e.g.][]{hopk06merge, some08bhev}, and some recent
studies have found enhanced evidence for mergers and interactions in
typical low-redshift AGNs \citep[e.g.,][]{koss10batagn,
  saba13agnenv}. However, a number of other studies of higher-redshift
sources have found that the hosts of typical moderate-luminosity AGNs are
no more likely than comparable ``normal'' galaxies to show
morphological disturbances characteristic of mergers
\citep[e.g.,][]{cist11agn, scha11xmorph, civa12ccosmos, koce12xmorph},
with little dependence on the merger fraction $f_{\rm merge}$ with
observed $L_{\rm AGN}$ \citep{koce12xmorph}. Other results have suggested that $f_{\rm
  merge}$ increases at very high $L_{\rm AGN}$
\citep[e.g.,][]{urru08qsohost} leading to the conclusion that black
hole growth can be triggered by major interactions, but only at the
highest luminosities \citep{scha12qsomorph,trei12merge}.

 In contrast to AGNs, star-forming galaxies consistently show a
 significant increase in the merger fraction across a wide range
 $L_{\rm IR}$ \citep[e.g.,][]{shi09irmorph,
   kart10irmorph,kart12irmorph}, suggesting that more rapidly
 star-forming galaxies are increasingly more likely to be associated
 with mergers (keeping in mind that being {\em associated} with
 mergers does not necessarily imply merger {\em triggering};
 \citealt{hopk10sfmerge}). Adopting the assumptions of our simple
 model, we can calculate the fraction of AGNs in mergers as a function
 of observed $L_{\rm AGN}$ for a scenario in which the merger fraction
 increases with SFR, and thus the time-averaged BH growth rate.  We
 adopt the relationship between $f_{\rm merge}$ and $L_{\rm IR}$
 obtained using deep {\em Hubble Space Telescope} rest-frame optical
 imaging of luminous star-forming galaxies \citep{kart12irmorph}. Here
 we use the fractions shown in Figure 9 of \citet{kart12irmorph}, for
 two different merger classifications. Over the range $11 <
 \log{(L_{\rm IR}\; [L_\odot]}) < 12.5$, the value of $f_{\rm merge}$
 for ``mergers and interactions'' increases from $\approx$\,0.3 to
 $\approx$\,0.6, while for ``mergers, interactions, and irregulars''
 (which may include some minor mergers), $f_{\rm merge}$ goes from
 $\approx$\,0.4 to $\approx$\,0.9. We linearly extrapolate these
 fractions to higher and lower $\log L_{\rm IR}$ (keeping the minimum
 and maximum fractions to 0 and 1), and assume that this relation
 between $f_{\rm merge}$ and $L_{\rm IR}$ is constant with redshift.
 This assumption is motivated by observations showing that the merger
 fraction for luminous infrared galaxies is similar in the local Universe
 \citep[e.g.,][]{wang06lirg, elli13irpair} to that at higher redshifts
 \citep[e.g.,][]{melb05irmorph, hung13irmorph}, as well as studies
 that directly measure  merger fractions with a
 full accounting for redshift effects, showing little evidence for
 variation with redshift \citep{shi09irmorph, kart12irmorph}. To obtain predictions for $f_{\rm merge}$ from our
 model, we determine a merger probability for each of our model
 galaxies based on its $L_{\rm IR}$ and the corresponding $f_{\rm
   merge}$ from \citet{kart12irmorph}. We then select model galaxies
 in bins of (instantaneous) $L_{\rm AGN}$ and calculate the total
 $f_{\rm merge}$ by averaging the merger probability for all objects
 in that bin.

\begin{figure}
\epsscale{1.15}
\plotone{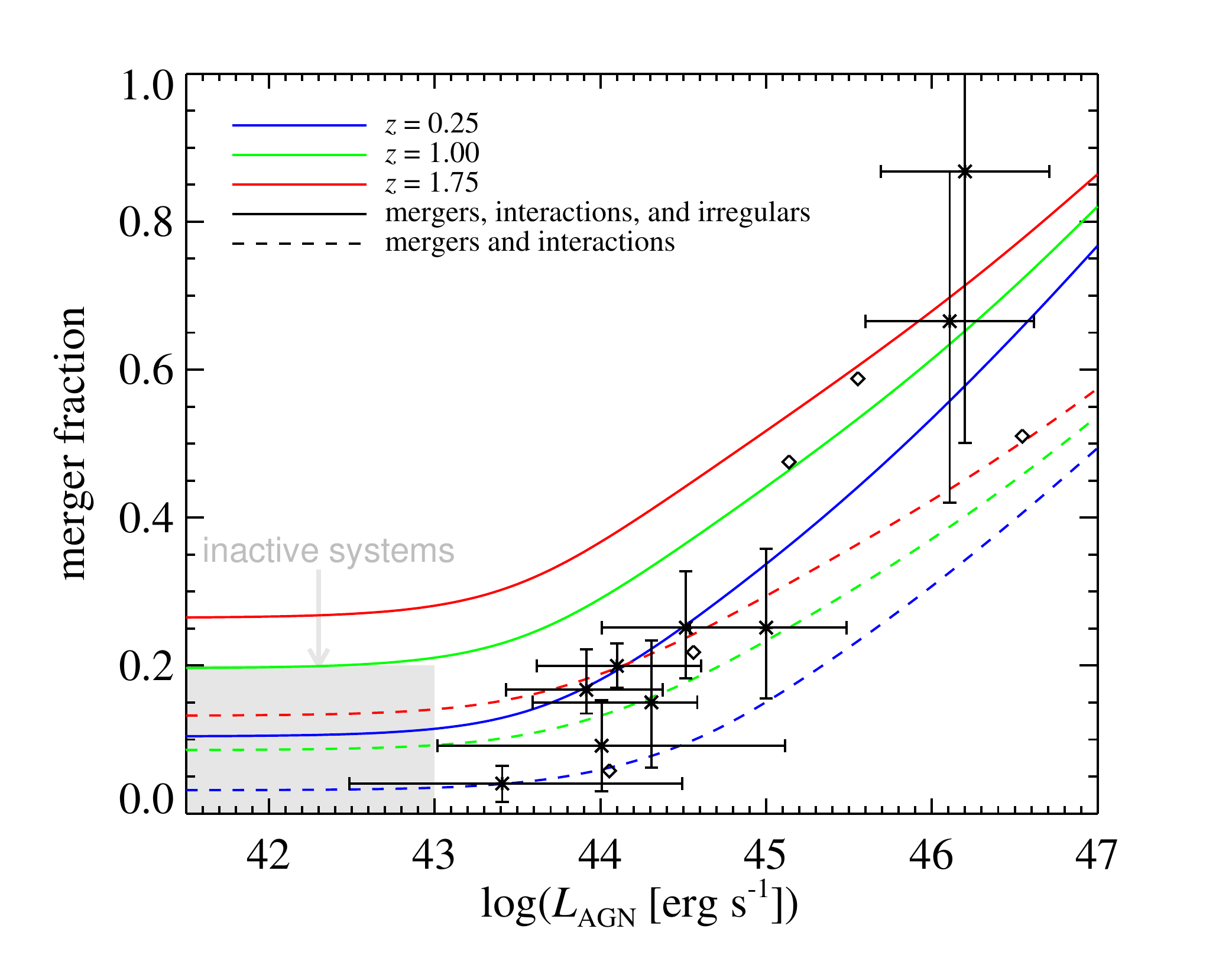}
\caption{Predictions of our fiducial model on the fraction of AGNs in
  mergers as a function of AGN luminosity. Data points are taken from
  the compilation of \citet{trei12merge}, and the gray shaded area
  indicates the typical range of merger fractions for inactive
  galaxies in the control samples studied by \citet{cist11agn} and
  \citet{koce12xmorph}. The colored curves show the predictions of our
  model assuming a correlation between merger fraction and $L_{\rm
    IR}$ determined by \citet{kart12irmorph}. The models are evaluated
  using the relationships from \citet{kart12irmorph} for ``mergers,
  interactions, and irregulars'' (solid lines) and ``mergers and
  interactions'' (dashed lines). The colors represent the model predictions for different
  redshifts. The model predicts a weak correlation between $f_{\rm
    merge}$ and $L_{\rm AGN}$ at low luminosities, with a stronger
  correlation emerging for the highest-luminosity systems, matching the
  general trends in the observational data. \label{fig:merge}}
\end{figure}

The model predictions are shown in Figure~\ref{fig:merge}, compared to
observational data compiled by \citet{trei12merge} and typical merger
fractions for control samples of inactive galaxies \citep{cist11agn,
  koce12xmorph}. We note that these observed AGN merger fractions are
obtained using a variety of different methods, such as detailed
morphological analysis and counts of close companions, which
complicates the comparison between individual data points (for a
discussion see \S\,1 of \citealt{kart12irmorph}). A clear trend is
nonetheless evident in the data, showing a weak dependence of $f_{\rm merge}$ on \lagn\ at low luminosity, with a strong upturn at high $L_{\rm
  AGN}$. The models clearly reproduce this trend, owing to the fact
that in our simple model, luminous AGNs are always associated with
rapidly star-forming galaxies while less luminous AGNs are drawn from a mix
of galaxy populations.  The weak dependence of the merger fraction on AGN
luminosity at low $L_{\rm AGN}$ may explain the observations that at
moderate to high redshift, moderate-luminosity AGNs have
indistinguishable merger fractions from normal galaxies
\citep{cist11agn, scha11xmorph, koce12xmorph}. As is clear from
Figure~\ref{fig:merge}, the merger fractions predicted by the model
depend on the choice of merger classification adopted from
\citet{kart12irmorph}; the inclusion of irregular systems naturally
produces a higher $f_{\rm merge}$ at all \lir. However despite these
differences, the general trends in the relationship between $f_{\rm
  merge}$ and \lagn\ are identical in the two cases, and match those
seen in the observational results.

Based on the assumptions of our simple model, we can determine the
fraction of the total BH growth that is associated with mergers at
different redshifts.  We use the curves of $f_{\rm merge}(L_{\rm
  AGN})$ shown in Figure~\ref{fig:merge} and the bolometric AGN LF
determined by \citet{hopk07qlf}.  Assuming a constant radiative
efficiency (as we have done throughout) such that the accretion rate
$\dot{m_{\rm BH}} \propto L_{\rm AGN}$, we compute the average $f_{\rm merge}$
for AGN weighted by the distribution in AGN growth rates, given by
$L_{\rm AGN} \phi(L_{\rm AGN})$. This yields a ``total'' merger
fraction of 23\% at $z=0$, rising to 72\% at $z=2$, for our model
including irregular systems (these fractions are 10\% and 44\%,
respectively, if we assume the fractions for only ``mergers and
interactions'').  These results are consistent with a picture in which
mergers are an important driver for global BH growth at high redshift,
with secular processes becoming increasingly dominant at low redshift
\citep[e.g.,][]{drap12agnmerge}.

\subsection{Effect of changing the accretion rate distribution}
\label{sec:distribution}

\begin{figure*}
\epsscale{1.1}
\plottwo{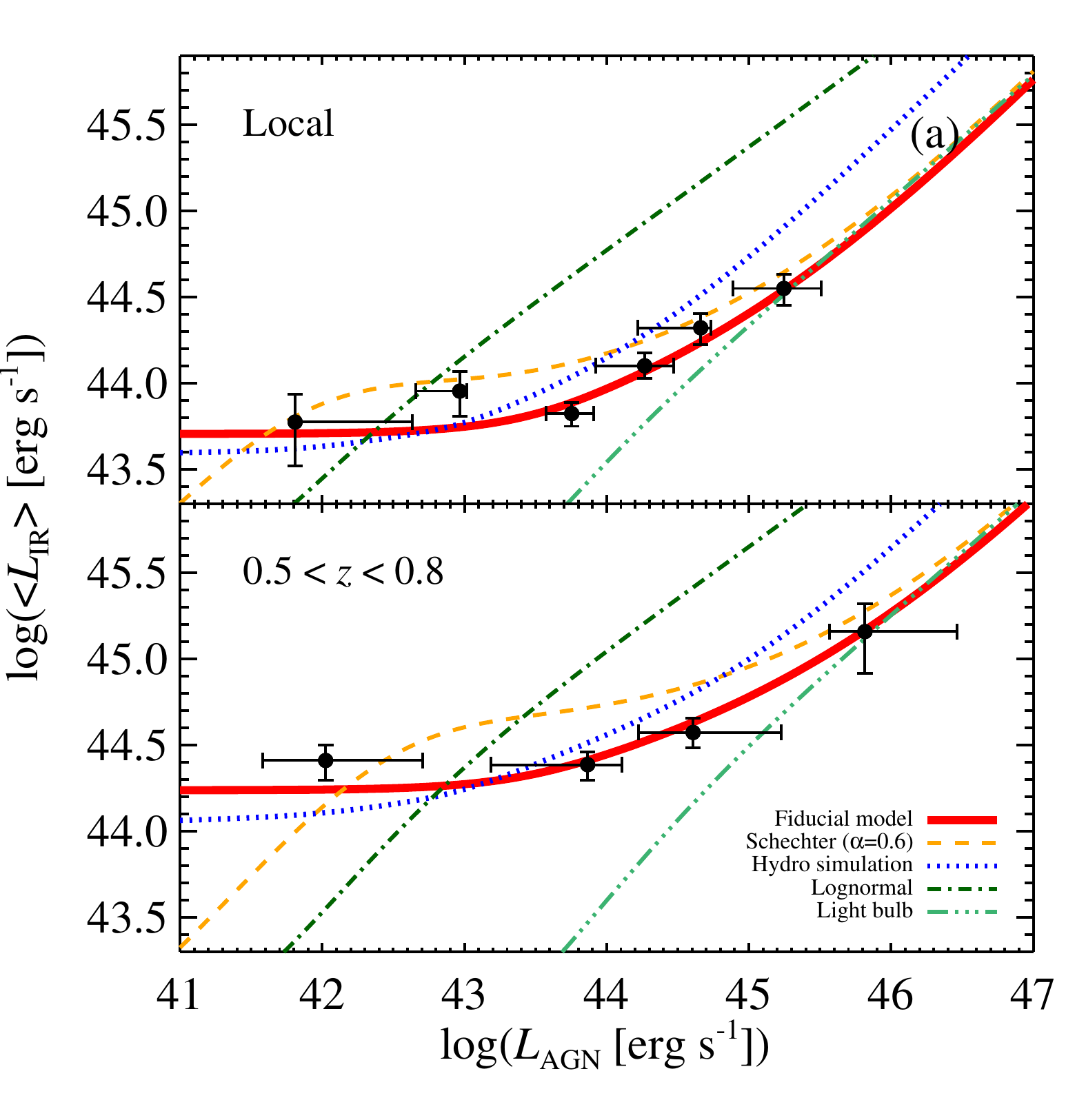}{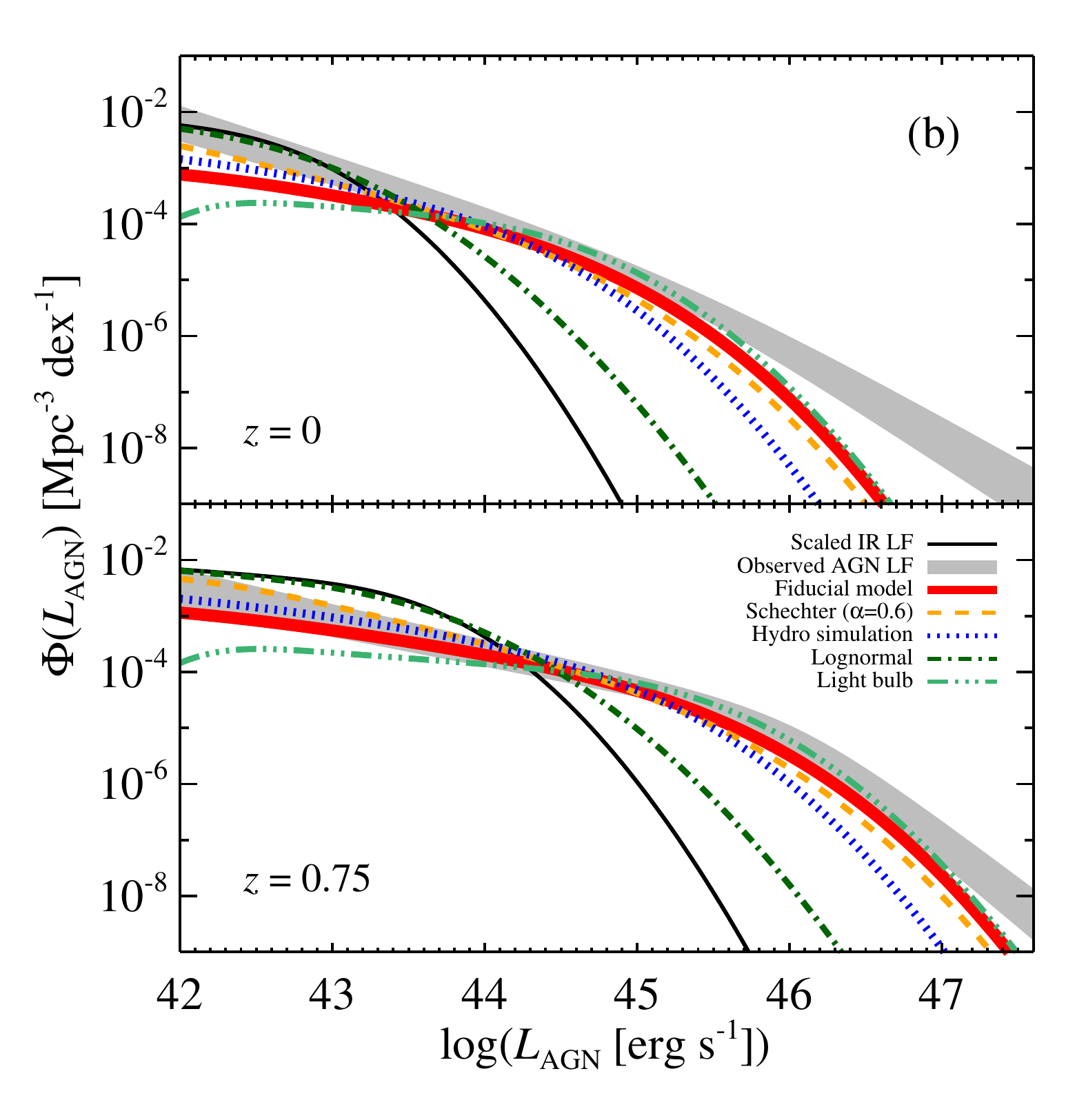}
\caption{Comparison of the predictions of different AGN luminosity
  distributions for ({\em left}) the average SFR as a function of
  $L_{\rm AGN}$, as in Figure~\ref{fig:agnvsf}(b), and ({\em right})
  the AGN LF, as in Figure~\ref{fig:lf}, for two representative
  redshift ranges. Both the distribution obtained by the
  \citet{nova11bhsim} simulation (blue dotted line) and our fiducial
  model (red solid line) can broadly reproduce both the observed
  trends with SF and the AGN LF. The steep observed AGN luminosity
  distribution at fixed stellar or BH mass \citep[modeled by the
    Schechter function with $\alpha=0.6$]{hopk09bulb, aird12agn} also
    produces a weak trend in the $\average{\rm SFR}$ versus $L_{\rm
      AGN}$ (although less closely matching the data) and fits the AGN
    LF particularly well. In our simple picture, a ``lognormal''
    luminosity distribution \citep{kauf09modes} yields too strong a
    correlation between $\average{\rm SFR}$ and $L_{\rm AGN}$ and
    fails to produce the high-luminosity tail of the AGN LF.
 \label{fig:dists}}
\end{figure*}

The previous analyses have focused on the predictions of our simple
model including our ``fiducial'' distribution of AGN accretion rates
(and equivalently, luminosities). We stress that while the shape of
our fiducial distribution is characteristic of those obtained from
recent theoretical and observational studies (as discussed in
\S\,\ref{sec:variability}), it is not obtained by a formal fit to the
data. Here we explore the implications of other distributions in the
accretion rate, focusing on the recent measurements and theoretical
results illustrated in Figure~\ref{fig:bhar}.

We first note that, by design, all the models predict the simple
linear relation between $\average{L_{\rm AGN}}$ and \lir\ shown in
Figure~\ref{fig:agnvsf}(a), independent of redshift and the choice of
accretion rate distribution. In contrast, the models produce
significant differences in the inverse relationship, between observed
$L_{\rm AGN}$ and $\average{L_{\rm IR}}$, as discussed in
\S\,\ref{sec:sfagn}. The model predictions for the different
luminosity distributions are shown in Figure~\ref{fig:dists}(a), along
with the observational data from \citet{rosa12agnsf}. The general
trends in the data are reproduced by all the models with
low-luminosity power-law distributions and a large dynamic range in
accretion rate. These all show a weak correlation between \lagn\ and
$\average{L_{\rm IR}}$ at moderate \lagn, with a stronger correlation
at high luminosity. The observed distribution at fixed mass (with
$\alpha=0.6$) further shows strong correlation at low \lagn, due to
the fact that the distribution is forced to cut off at a relatively
high average luminosity to avoid diverging (Figure~\ref{fig:bhar}), so
that very low-luminosity AGN are at the very bottom end of the
accretion rate distribution.  In contrast to the results for
distributions with a wide dynamic range, a relatively tight lognormal
distribution in accretion rates (either in the lognormal or ``light
bulb'' cases) yields a strong correlation between \lagn\ and
$\average{L_{\rm IR}}$ at all \lagn, in conflict with the
observations. We note that the ``light bulb'' model reproduces well
the strong correlation between $L_{\rm AGN}$ and $\average{L_{\rm
    IR}}$ at high luminosities, but does not produce the observed weak
correlation at moderate to low $L_{\rm AGN}$. This figure demonstrates
that models in which AGN experience a broad dynamic range in accretion
rate (and luminosity) can fit the general observed trends, and that
our fiducial distribution can reproduce the observations
particularly well.

We next focus on the predicted AGN LFs, as discussed in
\S\,\ref{sec:lf}. The model predictions for the different luminosity
distributions are shown in Figure~\ref{fig:dists}(b). Again, we find
that all the models with broad power-law luminosity distributions can
reproduce the general trends in the observed AGN LF, although they do
not produce enough AGNs at the very highest luminosities. (This may be
evidence for a somewhat flatter tail in the accretion rate
distribution than is modeled by a Schechter function; e.g.,
\citealt{aird13agn}.) In contrast, the lognormal distribution strongly
underpredicts the number of AGNs at moderate to high luminosities,
since the AGN luminosity is fairly tightly tied to
\lir. Interestingly, the ``light bulb'' model results in a similar LF
to that predicted by our fiducial model on the high-luminosity end,
although it produces fewer low-luminosity AGNs. This general agreement
with the observed AGN LF has been found by previous studies of ``light
bulb'' models where the accretion rates are scaled to BH or galaxy mass rather than SFR \citep[e.g.,][]{siem97qsolf, conr13qso}. We conclude that, in
the context of our AGN variability model, reproducing the general
trends in the observed AGN LF requires that the AGN luminosity
distribution must extend to relatively high $L_{\rm AGN}$ above the
long-term average luminosity.  However, predictions for the AGN LF are
relatively insensitive to the precise choice of luminosity
distribution, indicating that the relationship between
$\average{L_{\rm IR}}$ and \lagn\ may provide a better constraint on
the nature of the AGN variability.

\section{Discussion}
\label{sec:dis}

We have shown that the observed relationships between AGN luminosity,
SF, and galaxy mergers, as well as the relative shapes of the IR and
AGN LFs, can be broadly explained by a simple picture in which BH
accretion rates are perfectly connected to SFRs, but subject to
short-timescale variability over a large dynamic range. This picture
may have significant implications for studies of AGN triggering, as it
implies that the observed {\em instantaneous} luminosity of an AGN is
a weak indicator of the {\em average} BH accretion rate on the
timescales of the galaxy evolution processes that may be expected to
drive the long-term growth of BHs. Thus powerful quasars may represent
brief upward fluctuations in the AGN luminosity of otherwise passive
systems, while seemingly ``normal'' galaxies may have experienced
powerful AGN activity and rapid BH growth in the recent past.

AGN feedback is not explicitly included in this analysis, however the
strong correlation between SFR and long-term BH accretion rate
prescribed by our model may suggest indirectly that some feedback
processes are occuring. Small-scale feedback, in which the energy
released by the BH limits its own gas supply, is consistent with our
model as it is the key physical process that drives rapid variability
of the AGN over a large dynamic range in a number of theoretical
studies \citep[e.g.,][]{hopk05life1, ciot10flare, nova11bhsim,
  gabo13bhsim}. On larger scales, a particularly tight connection
between SFR and BH accretion is predicted by some models of positive
feedback in which AGN activity triggers SF
\citep[e.g.,][]{zubo13agnsb, naya13feed}, but such a correlation over long
timescales can arise in models with zero or even negative AGN feedback
\citep[e.g.,][]{angl13bhsim, gabo13bhsim}, so these results alone do
not enable us to draw any strong conclusions about the
effects of BH feedback on galaxy-wide star formation.

Indeed, despite its remarkable success in reproducing a range
of observational results, our model is too simplistic to yield
information on the details of AGN fueling and variability.  For
example, our model assumes a perfect proportionality between SFR and
long-term BH accretion rate and does not allow for any scatter in this
relationship. Relatively small scatter in the BHAR/SFR ratio would be
equivalent to simply broadening the observed AGN luminosity
distribution at a fixed SFR, although large intrinsic scatter in this
ratio would flatten the observed correlation between $\average{L_{\rm
    AGN}}$ and \lir\ shown in Figure~\ref{fig:agnvsf}(a), and so would
be inconsistent with observations \citep[for a discussion
  see][]{chen13agnsf}. While the strong observed correlation
suggests that the intrinsic scatter in the BHAR/SFR ratio is
relatively small, this scatter must be constrained independently in
order to extract the true variability in AGN luminosities.

Another limitation of our model is that, in order to keep it as
simple as possible, we do not include any consideration of galaxy or
BH masses. We therefore explicitly ignore the dependence of the
Eddington limit on BH mass. This may thus cause us to overpredict the
number of luminous AGN in small but rapidly star-forming galaxies with
small BHs, and underpredict the maximum luminosities of massive
galaxies with large BHs. Indeed, the tendency of AGNs to be found in
relatively massive galaxies \citep[e.g.,][]{kauf03host, colb05xhost,
  hagg10champ, xue10xhost, card10xhost} and halos
\citep[e.g.,][]{hick09corr, hick11qsoclust, coil09xclust, star11xclust, alle11xclust,
  capp12xclust}, and the corresponding existence of relatively
powerful AGNs in massive but passive systems (as discussed in
\S\,\ref{sec:lf}) are likely a direct consequence of the Eddington
limit \citep[e.g.,][]{hopk09lowlum, aird12agn}. In the context of our model,
introducing an Eddington limit would be equivalent to varying the
luminosity at which the distributions cut off on the high end,
depending on the relationship between SFR and BH mass in each galaxy.

A more sophisticated version of our model would therefore account for
the joint distribution of BH (and galaxy) masses and SFRs, while
including an explicit Eddington limit. This could be achieved
analytically by a similar process to that described here, but expanded
to include a careful treatment of the observed redshift evolution in
the galaxy stellar mass and luminosity functions (similar to the
analysis of \citealt{conr13qso}) while also accounting for the
distribution of SFRs as a function of galaxy mass, as in recent
studies of galaxy formation \citep[e.g.][]{peng10sfgal, behr13sfgal,
  lill13sfgal}.  Alternatively, our AGN variability prescriptions
could be incorporated into semi-analytic models of galaxy formation
based on dark matter halo merger trees, which explicitly track the
stellar and BH masses and SFRs of each component galaxy
\citep[e.g.][]{bowe06gal, some08bhev, fani12agn, fani13xclust}. By
comparing these more sophisticated models to observations, we may be
able to obtain a reliable picture of the variability of AGNs and the
connection between BH accretion and SF.

As discussed in \S\,\ref{sec:intro}, a complete understanding of AGN
variability may potentially reconcile a range of seemingly
contradictory observations about the relationships between AGNs and
their host galaxies.  However, the stochastic nature of the
variability also requires that we employ an inherently statistical
approach in observational studies by measuring the {\em distribution}
in AGN accretion rates as a function of galaxy properties. Currently,
X-ray surveys provide one of the most robust methods for probing AGN
accretion over a wide range of Eddington ratios, host galaxy
properties, and redshifts \citep[e.g.,][]{hopk09lowlum}, and have
enabled the first such statistical studies at moderate to high
redshift \citep[e.g.,][]{hick09corr, aird12agn, bong12xgal}.
  Using these techniques and existing X-ray and far-IR observations,
  it might be possible to obtain a measurement of the distribution of
  $L_{\rm AGN}$ in broad bins of SFR, which would provide an
  interesting constraint on our model's prediction\footnotemark\ that
  the distribution in $L_{\rm AGN}/{\rm SFR}$ is independent of SFR or
  redshift.  An alternative could be to measure the distribution of
  $L_{\rm IR}$ in bins of $L_{\rm AGN}$, for which the model also
  makes clear predictions as shown in
  Figure~\ref{fig:sfdist}. However, to explore the distribution of AGN
  accretion rate as a function of several interesting host galaxy
  properties (i.e., SFR, stellar mass, redshift) would require larger
  X-ray AGN samples than are currently available. Our results
therefore provide motivation for future deep, wide extragalactic
surveys that will obtain large samples of AGNs over a wide range in
redshift and luminosity.

\footnotetext{Data and software for computing the predictions of the model at a given redshift and $L_{\rm AGN}$ or $L_{\rm IR}$ are available at http://www.dartmouth.edu/\string~hickox/sfagn.php}

\begin{acknowledgements}

We thank David Rosario, Sara Ellison, and Philip Hopkins for helpful
discussions, and are grateful to the anonymous referee for
  constructive comments that improved the paper. J.R.M.\ and
D.M.A.\ acknowledge generous support from the Leverhulme Trust.
F.C.\ acknowledges support by the NASA contract 11-ADAP11-0218.
C.-T.J.C.\ acknowledges support from a Dartmouth Fellowship. This work
was supported in part by {\em Chandra} grant SP8-9001X. This research
has made use of NASA's Astrophysics Data System.

\end{acknowledgements}

\bibliographystyle{apj} \bibliography{research}
\end{document}